\documentclass{elsart}
\usepackage{graphicx,epsfig,amssymb,afterpage}
\newcommand\textrmn[1]{\textrm{\scriptsize #1}}
\journal{Physics Letters B}
\begin{document}
\begin{frontmatter}

\title{Measurement of the Ratio of the Vector to Pseudoscalar 
Charm Semileptonic Decay Rate  ${{\Gamma(D^+\to \overline{K}^{*0}\mu^+\nu_\mu)}\over
{\Gamma(D^+\to \overline{K}^0\mu^+\nu_\mu)}}$}

\subtitle{\large\rm{The FOCUS Collaboration\thanksref{focu}}}
\thanks[focu]{\mbox{See 
http://www-focus.fnal.gov/authors.html for additional author information.}}
\addtocounter{corauth}{1}
\author[ucd]{J.~M.~Link}
\author[ucd]{P.~M.~Yager}
\author[cbpf]{J.~C.~Anjos}
\author[cbpf]{I.~Bediaga}
\author[cbpf]{C.~G\"obel}
\author[cbpf]{A.~A.~Machado}
\author[cbpf]{J.~Magnin}
\author[cbpf]{A.~Massafferri}
\author[cbpf]{J.~M.~de~Miranda}
\author[cbpf]{I.~M.~Pepe}
% ...
\corauth[cor]{Corresponding author; present address:
Vanderbilt University, Dept. of Physics and Astronomy, 6301 Stevenson Ctr.,
Nashville TN, 37235 (USA). Tel.: 615-343-8295 (Cell: 303-579-6032). 
{\it{E-mail}}: will.johns{@}vanderbilt.edu}
\author[vu]{E.~Luiggi}
\author[vu]{M.~Nehring}
\author[vu]{P.~D.~Sheldon}
\author[vu]{E.~W.~Vaandering}
\author[vu]{M.~Webster}
\author[wisc]{M.~Sheaff}
\address[ucd]{University of California, Davis, CA 95616}
\address[cbpf]{Centro Brasileiro de Pesquisas F\'isicas, Rio de Janeiro, RJ, Brasil}
\address[cinv]{CINVESTAV, 07000 M\'exico City, DF, Mexico}
\address[cu]{University of Colorado, Boulder, CO 80309}
\address[fnal]{Fermi National Accelerator Laboratory, Batavia, IL 60510}
\address[fras]{Laboratori Nazionali di Frascati dell'INFN, Frascati, Italy I-00044}
\address[ugj]{University of Guanajuato, 37150 Leon, Guanajuato, Mexico}
\address[ui]{University of Illinois, Urbana-Champaign, IL 61801}
\address[iu]{Indiana University, Bloomington, IN 47405}
\address[korea]{Korea University, Seoul, Korea 136-701}
\address[kp]{Kyungpook National University, Taegu, Korea 702-701}
\address[milan]{INFN and University of Milano, Milano, Italy}
\address[nc]{University of North Carolina, Asheville, NC 28804}
\address[pavia]{Dipartimento di Fisica Nucleare e Teorica and INFN, Pavia, Italy}
\address[pr]{University of Puerto Rico, Mayaguez, PR 00681}
\address[sc]{University of South Carolina, Columbia, SC 29208}
\address[ut]{University of Tennessee, Knoxville, TN 37996}
\address[vu]{Vanderbilt University, Nashville, TN 37235}
\address[wisc]{University of Wisconsin, Madison, WI 53706}

\begin{abstract}
Using a high statistics sample of photo-produced charm particles from the FOCUS
experiment at Fermilab, we report on the measurement of 
the ratio of semileptonic
rates ${{\Gamma(D^+\to {K}\pi\mu^+\nu_\mu)}\over
{\Gamma(D^+\to \overline{K}^0\mu^+\nu_\mu)}}=0.625\pm0.045\pm0.034$. Allowing
for the $K\pi$ 
S-wave interference measured in \cite{Link:2002wg}, we extract the vector to pseudoscalar 
ratio  ${{\Gamma(D^+\to \overline{K}^{*0}\mu^+\nu_\mu)}\over
{\Gamma(D^+\to \overline{K}^0\mu^+\nu_\mu)}}=0.594\pm0.043\pm0.033$ and the ratio
${{\Gamma(D^+\to \overline{K}^{0}\mu^+\nu_\mu)}\over
{\Gamma(D^+\to K^-\pi^+\pi^+)}}=1.019\pm0.076\pm0.065$. Our results show a lower 
ratio for ${{\Gamma(D\to K^*\ell\nu})\over{\Gamma(D\to K\ell\nu)}}$ than has been 
reported recently and indicate the current world average
branching fractions for the decays $D^+\to \overline{K}^{0}(\mu^+,e^+)\nu_{\mu,e}$ are low.
Using the world average ${\textrm{B}}(D^+\to K^-\pi^+\pi^+)$ \cite{Hagiwara:fs} we 
extract $\textrm{B}(D^+\to \overline{K}^0\mu^+\nu)=(9.27\pm0.69\pm0.59\pm0.61)~\%.
$
\end{abstract}
% Text of abstract
\end{frontmatter}

\addtocounter{footnote}{-1}

\section{Introduction}

There is currently some controversy concerning the relative rates of
the charm vector semileptonic decays that proceed via a $K^*$ and the
charm pseudoscalar semileptonic decays that proceed via a $K$. Theorists
originally expected the rates to be about the same \cite{Wirbel:1985ji}, 
but other theoretical predictions and experimental 
measurements in the 90's tend to favor a smaller vector 
semileptonic rate (see Tables \ref{tab:exp} and \ref{tab:theos}). 

A more recent experimental result \cite{Brandenburg:2002eu} indicates that the
ratio of rates is closer to unity than measured previously. Rather than measure the 
rate for the pseudoscalar and vector semileptonic decays directly, as was done in
previous measurements by the same experiment \cite{Crawford:1991zd,Bean:zv},
the result uses the average from the PDG for the determination of
$\textrm{B}(D^+\to K^0e^+\nu_e)$ \cite{Hagiwara:fs}. Other experiments
measure the semileptonic {\it{rates}} for $D^+$ and $D^0$ decays and
form a ratio of vector to pseudoscalar rates (see Table \ref{tab:exp}). 

Since the $D^+$ and the $D^0$ particles only differ by the light quark,
exclusive semileptonic rates for the decays of these particles are expected 
to be equal through SU(3) symmetry. A comparison using the current world 
averages of the pseudoscalar decay branching fractions along with the 
$D^+$ and $D^0$ lifetimes \cite{Hagiwara:fs} 
indicates that the pseudoscalar semi-electronic rates (the error on the
$D^+$ pseudoscalar semi-muonic rate is too large for a meaningful comparison)
are different at the 99\%  confidence level: $\Gamma(D^+\to \overline{K}^0e^+\nu_e)-\Gamma(D^0\to
K^-e^+\nu_e)=-25\pm9.7~\textrm{ns}^{-1}$. 
This result is surprising and merits further investigation.
We intend to show in this paper that the difference in rates and the
recent CLEO \cite{Brandenburg:2002eu} result are in part due to the pseudoscalar semileptonic branching
fraction reported in the PDG \cite{Hagiwara:fs} for the decay $D^+\to K^0\ell^+\nu_\ell$ being too low. 

We measure the ratio 
${{\Gamma(D^+\to {K}\pi\mu^+\nu_\mu)}\over{\Gamma(D^+\to \overline{K}^0\mu^+\nu_\mu)}}$ 
directly, using decays with a very similar topology. Previous measurements of this ratio 
used comparisons between the $D^+$ and the $D^0$, relied on PDG
branching ratios, and/or used decays with different topologies (for instance where one of the modes
requires an added pion). By reconstructing the
neutral kaon in the microvertex detector of the FOCUS experiment \cite{Frabetti:1990au,Link:2002zg} through the
decay $K^0_S\to\pi^+\pi^-$, we take advantage of the studies and
work performed to produce precise lifetime measurements of the
long-lived charm particles. By measuring the $D^+$ decay, we take advantage
of the extensive work performed to understand the decay $D^+\to {K}^-\pi^+\mu^+\nu_\mu$ 
\cite{Link:2002wg,Link:2002ev,Link:2002nj}
and our result is the first to incorporate the interference effects described in \cite{Link:2002ev} and
measured in \cite{Link:2002wg}.

The data for this analysis
were collected using the Wideband photoproduction experiment FOCUS during the 
1996--1997 fixed-target run at Fermilab. The FOCUS detector is a large aperture, 
fixed-target spectrometer with excellent vertexing and particle identification
used to measure the interactions of high energy photons on a segmented BeO 
target. The FOCUS beamline \cite{Frabetti:1992bn} and 
detector \cite{Frabetti:1990au,Link:2002zg,Link:2002ev,Link:2001pg} have 
been described elsewhere.

\begin{table}[htb]
\begin{minipage}{\textwidth}
\caption{Previous results compared to the FOCUS result. Notice that some results
\cite{Anjos:1988ue,Frabetti:1993vz,Frabetti:1995xq} are admixtures of different 
charm species related through the isospin argument, \cite{Frabetti:1993vz,Frabetti:1995xq}
are correlated since the same $D^+\to \overline{K}^{*0}\mu^+\nu$ result is used, and
in the CLEO(02) result \cite{Brandenburg:2002eu}, the $\Gamma(D^+\to \overline{K}^0e^+\nu)/\Gamma_\textrmn{Total}$ comes from the
PDG00 \cite{PDG00:2000sj} estimate for  $\Gamma(D^+\to \overline{K}^0\ell^+\nu)/\Gamma_\textrmn{Total}$.
}
\label{tab:exp}
\begin{center}
\begin{tabular}{cccccl} \hline 
Experiment& Quantity & Result \\ \hline 
CLEO(91)\cite{Crawford:1991zd} & ${{\Gamma(D^0\to {K}^{*-}e^+\nu)}\over{\Gamma(D^0\to K^-e^+\nu)}}$                 & $0.51\pm 0.18\pm0.06$       \\ \hline 
CLEO(93)\cite{Bean:zv}  & ${{\Gamma(D^0\to {K}^{*-}e^+\nu)}\over{\Gamma(D^0\to K^-e^+\nu)}}$                        & $0.60\pm 0.09\pm0.07$       \\ \hline 
CLEO(93)\cite{Bean:zv}  & ${{\Gamma(D^+\to \overline{K}^{*0}e^+\nu)}\over{\Gamma(D^+\to \overline{K}^0e^+\nu)}}$               & $0.65\pm 0.09\pm0.10$       \\ \hline 
E691(89)\cite{Anjos:1988ue} & ${{\Gamma(D^+\to \overline{K}^{*0}e^+\nu)}\over{\Gamma(D^0\to K^-e^+\nu)}}$           & $0.55\pm 0.14$              \\ \hline 
E687(93)\cite{Frabetti:1993vz}    & ${{\Gamma(D^+\to \overline{K}^{*0}\mu^+\nu)}\over{\Gamma(D^0\to K^-\mu^+\nu)}}$ & $0.59\pm0 .10\pm0.13$       \\ \hline     
E687(95)\cite{Frabetti:1995xq}  & ${{\Gamma(D^+\to \overline{K}^{*0}\mu^+\nu)}\over{\Gamma(D^0\to K^-\mu^+\nu)}}$   & $0.62\pm0.07\pm0.09$        \\ \hline	   
CLEO(02)\cite{Brandenburg:2002eu}  & ${{\Gamma(D^+\to \overline{K}^{*0}e^+\nu)}\over{\Gamma(D^+\to \overline{K}^0e^+\nu)}}$    &
$0.99\pm0.06\pm0.07\pm0.06~(\pm0.12)$\footnote{The PDG00 \cite{PDG00:2000sj} error for $\Gamma(D^+\to \overline{K}^0\ell^+\nu)/\Gamma_{Total}$, omitted \cite{Cleo:Yuichi}
in the CLEO \cite{Brandenburg:2002eu} result, is shown in parentheses.} \\ \hline
FOCUS(04) & ${{\Gamma(D^+\to \overline{K}^{*0}\mu^+\nu)}\over{\Gamma(D^+\to \overline{K}^0\mu^+\nu)}}$   &                       $0.594\pm0.043\pm0.030$     \\ \hline
\end{tabular}
\end{center}
\end{minipage}
\end{table}
\addtocounter{footnote}{1}

\begin{table}[ht]
\caption{Theoretical predictions for the Vector to Pseudoscalar ratio. No distinction
is made for electrons, muons or charm species. In some cases, listed with a (*), the ratio was calculated from
information given in the publication.}
\label{tab:theos}
\begin{center}
\begin{tabular}{cccl} \hline 
Model                             & ${{\Gamma(D\to {K}^{*}\ell^+\nu)}\over{\Gamma(D\to K\ell^+\nu)}}$ \\ \hline 
WSB(85*) \cite{Wirbel:1985ji}       & $1.13$                  \\ \hline
LFR(90*) \cite{Jaus:au}             & $0.7,0.67$              \\ \hline
SUMR(91)\cite{Ball:bs}             & $0.50\pm0.15$           \\ \hline
LAT(92*) \cite{Lubicz:1991bi}       & $0.86\pm0.27$           \\ \hline
LAGR(93*) \cite{Casalbuoni:1992dx}  & $0.56$                  \\ \hline
LAT(94)    \cite{Abada:1993dh}      & $1.1\pm0.6\pm0.3$       \\ \hline
MITBAG(94) \cite{Sadzikowski:iy}     & $0.56$                  \\ \hline
ISGW2(95) \cite{Scora:1995ty}       & $0.54$                  \\ \hline
QPT(96)  \cite{Faustov:xe}          & $0.65$                  \\ \hline
RQM(96)  \cite{Jaus:np}             & $0.57$                  \\ \hline
QM(97) \cite{Melikhov:1996pr}      & $0.62$                  \\ \hline
LFM(97) \cite{Demchuk:1997uz}      & $0.68$                  \\ \hline
SR(97*)  \cite{Yang:ar}             & $0.47^{+0.19}_{-0.17}$  \\ \hline
QM(00),DISP(01) \cite{Melikhov:2000yu,Melikhov:2001zv}& $0.63$                  \\ \hline
COVQ(01*) \cite{Merten:2001er}      & $1.01,0.72$             \\ \hline
CVLFD(01) \cite{Nasriddinov:qh}    & $0.66,0.64,0.67$   \\ \hline
PCL(01) \cite{Kondratyuk:2000ka}   & $0.54,0.63,0.57,0.67$                  \\ \hline
EFT(02) \cite{Wang:2002zb}         & $0.5\pm0.2$             \\ \hline
\end{tabular}
\end{center}
\end{table}

\section{Event Selection}

We identify $D^+$'s through the
3-body decay  $D^{+}\to h^-\pi^+\mu^+\nu_\mu$ (where the
$h$ represents a pion or a kaon and charge conjugate modes are
implied throughout this paper). To search for candidate events,
a search strategy common to both modes is employed. This is possible
since roughly 10\% of all $K^0_S$'s reconstructed in the $\pi^+\pi^-$ 
channel are reconstructed using hits from the FOCUS silicon 
microvertex detectors.\footnotemark
\footnotetext{The technique used in \cite{Link:2001zj}
to reconstruct $D^{+}\to K^0_S\pi^+$ using the bulk of the 
$K^0_S$'s is not applicable in this case due to the missing neutrino.}
Where possible,
we have chosen cuts similar to those optimized in other FOCUS analyses to enhance
charm content and particle identification.

Two opposite sign tracks reconstructed using information from the silicon detectors
are required to form a vertex with 
a confidence level exceeding 1\%.
To suppress background from short-lived, primarily non-charm particles
produced in the targets,
this 2-track vertex is required to occur at least 1 standard deviation
outside of target material. 
The two tracks are then formed into a single track that is combined with
a candidate muon to make a putative $D^+$ vertex with a confidence level exceeding 1\%. 
This $D^+$ vertex is required to occur 
at least 1 standard deviation outside of target material as well.

Due to the relatively long lifetime and Lorentz boost
of charm candidates, the primary interaction vertex and secondary $D^+$ decay
vertex can have a significant separation along the beam direction.
Tracks not used to form a $D^+$ are used to construct a set of candidate primary
vertices with confidence level greater than 1\%. The highest multiplicity
primary vertex candidate with the highest separation, $\ell$, from the secondary
in units of error, $\sigma_\ell$, greater than 3 is retained. Since the 
significance of the separation between the interaction and decay vertices is  
an essential tool used to separate charm from background, we perform
a scan in this variable between $\ell/\sigma_\ell=3$ and $\ell/\sigma_\ell=23$ 
(roughly a factor of 3 in yield) to judge the stability of our result. 
We find that our results become stable after a cut of $\ell/\sigma_\ell>5$ is reached.
Additionally, the fit quality for the determination of the $D^+\to K^0_S\mu^+\nu_\mu$
yield becomes optimal ($\chi^2/\textrm{DOF}<1$) between  $\ell/\sigma_\ell=11$ and
$\ell/\sigma_\ell=19$. For the final sample, a value of $\ell/\sigma_\ell=13$ is used.
The ratios of the yields of individual modes to that expected from simulations 
exhibit the same stability. This is a strong indication that our cuts are effective
at removing any short-lived (non-charm) backgrounds that could mimic signal.

Muon candidates are required to be within the acceptance of the inner
muon system in FOCUS \cite{Link:2002ev}.
Since the rate of muon misidentification
increases at low momentum, we require that the momentum of muon
candidate tracks be greater than $13~\textrm{GeV/c}$. 
Muon candidates are required to have associated hits in 
the muon system sufficient to meet a minimum confidence level 
of 5\% for the muon hypothesis where at least 5 of 6 planes of the 
detector must record hits consistent with the candidate track.

To separate decays proceeding through the $K^-\pi^+\mu^+\nu_\mu$ channel, from those
decaying through the $K^0_S\mu^+\nu_\mu$ channel, we use additional vertex, particle
ID and invariant mass requirements. 

The vertex representing the $K^-\pi^+$ is required to occur within
3 standard deviations of the $(K^-\pi^+)\mu^+$ vertex, and a three track vertex
formed from the $K^-$,$\pi^+$ and $\mu^+$ tracks must exceed a confidence level of 5\%.
Since the $K^0_S$ lifetime is long compared even to the charm lifetime, we find
a very effective cut to reduce non-$K^0_S$ contamination is to impose a requirement
that the $K^0_S$ vertex have a large separation from the charm decay vertex. We 
require that the vertex representing the $K^0_S\to \pi^-\pi^+$ decay occur
at least 15 standard deviations downstream of the $(\pi^-\pi^+)\mu^+$
vertex.

We use the \v{C}erenkov system \cite{Link:2001pg} to identify pions and
kaons. 
For each track, $W_\textrmn{obs} = - 2 \log {\mathcal{(L)}}$ is computed, where
$\mathcal{L}$ is the likelihood that a track is consistent with a given
particle hypothesis. For the track in the $K^-\pi^+\mu^+$ vertex with charge
opposite to the muon, we require
$W_\textrmn{obs}(\pi)-W_\textrmn{obs}(K)$ (kaonicity) be greater than 2.0, and
for the track same charge as the muon in the $K^-\pi^+\mu^+$ vertex, we require
$W_\textrmn{obs}(K)-W_\textrmn{obs}(\pi)$ (pionicity) be greater than 0.0.
For pions that form a candidate $K^0_S$, we require that the pion likelihood
be favored over the particle hypothesis that forms the minimum likelihood
$W_{min}-W_\textrmn{obs}(\pi)$ be greater than -5.0 (a loose cut).

Background from $D^{+}\to K^-\pi^+\pi^+$ and $D^{+}\to K^0_S\pi^+$, where a pion is
misidentified as a muon, is reduced by requiring that the visible mass $M(h^-\pi^+\mu^+)<1.8~\textrm{GeV/c}^2$.
In order to suppress background from $D^{*+}\to D^0\pi^+ \to (K^-\mu^+\nu_\mu)\pi^+$
we require $M(K^-\pi^+\mu^+)-M(K^-\mu^+)>0.2$. In order to enrich the $K^0_S$ sample
we require the $K^0_S$ invariant mass be within 2 standard deviations of the nominal
reconstructed value. 

To reduce backgrounds from topologies consistent with extra tracks coming from the
secondary vertex (such as expected backgrounds like $D^0\to \overline{K}^0\pi^-\mu^+\nu_\mu$ 
or high multiplicity charm and non-charm backgrounds not present in the simulation) we reject candidates
where any track not used to create the secondary vertex can be included in the secondary
vertex with a vertex fit confidence level exceeding 
$10$\%.

\section{Analysis}

We fit the $K^0_S\mu^+$ invariant mass and the $K^-\pi^+$ invariant mass.
Fit components are a combination of Monte Carlo histograms: one generated with the mode of interest 
and others representing background components. A maximum likelihood is used where the predicted 
number of events in a bin$_i$ is described by:

$$N(\textrm{predicted})_i=P_S\textrm{Signal}_i+P_j\textrm{Background}_{i,j}$$

where $P_S$ and the $P_j$'s (more than one background shape is used) are fit parameters. 
The number of predicted signal events in the data is then described 
by $\sum_i^{\#\textrmn{bins}} P_S\textrm{Signal}_i$ 
where $\textrm{Signal}_i$ is the number of reconstucted Monte Carlo events 
for the mode of interest in a given bin. 

We find that the fit to determine the yield of the $D^+\to K^-\pi^+\mu^+\nu_\mu$ events 
requires only 2 components: one for signal and one for background. The 
$D^+\to K^-\pi^+\mu^+\nu_\mu$ matrix element has only recently been fully measured,
and we simulate the signal shape and efficiency with the results from \cite{Link:2002wg}.
We represent the background shape
in the $K^-\pi^+$ mass by generating a Monte Carlo in which we simulate all known charm
decay backgrounds while removing $D^+\to K^-\pi^+\mu^+\nu_\mu$ from the generated
particle mix. 

We have tried several approaches to fitting the invariant ${K}^{0}_S\mu^+$ mass histogram.
To generate $D^+\to \overline{K}^0\mu^+\nu_\mu$ events, we use the simple pole form for the 
pseudoscalar to pseudoscalar semileptonic form factor where:

$$f_\pm(q^{2}) = f_\pm(0) / {(1-{q^{2}/M_\textrmn{pole}^{2})}},~~q^{2}=(P_{D}-P_\textrmn{kaon})^2=M_\textrmn{W-virtual}^2.$$

We use $M_\textrmn{pole}=2.11 ~\textrm{GeV/c}^2$ and $f_-/f_+=-0.7$. Since we accept almost the entire 
${K}^{0}_S\mu^+$ invariant mass range, small changes in the choice of $M_\textrmn{pole}$ and
$f_-/f_+$ have a negligible effect on our final result. To represent the background
in the data, several techniques were investigated.  

If the background in the data is primarily due to $D\to \overline{K}^0\pi\mu^+\nu_\mu$
when a $\pi^0$ or $\pi^+$ is missed and $D^+\to \overline{K}^0\mu^+\nu_\mu$ when either 
the $K^0_S$ or the $\mu^+$ is not from the $D^+$, it should be sufficient
to perform a fit including only these two components. We find that this approach produces
consistent results but poor fit quality at low $\ell/\sigma_\ell$'s. Fit quality improves 
though at the cost of stability if the 2 (or more) lowest mass bins are removed from the fit. 
We also find a slight improvement in fit quality and stability at high $\ell/\sigma_\ell$ 
if a Breit-Wigner component centered at $892 ~ \textrm{MeV/c}^2$ with a width of $50~ \textrm{MeV/c}^2$ is 
added to the fit. 
 
In order to try and improve the fit quality, we added a background shape to the 
previous 3 component fit by generating a Monte Carlo in which we simulate all known charm
decay backgrounds while removing $D^+\to K^0_S\mu^+\nu_\mu$ from the generated
particle mix. We found that the fit quality did not improve, the error on the returned
fit increased, and the results were not stable below $\ell/\sigma=9$. Even though
the results agreed quite well at higher $\ell/\sigma$, we decided to investigate 
this behavior further.

%Both of these fit techniques have been tested (without the added  
%Breit-Wigner component) on simulated data using our best
%knowledge of charm photoproduction and decay branching fractions.
%We were able to faithfully reproduce the PDG \cite{Hagiwara:fs} ratio 
%${{\Gamma(D^+\to ({K}^-\pi^++K^0\pi^0)\mu^+\nu_\mu)}\over
%{\Gamma(D^+\to \overline{K}^0\mu^+\nu_\mu)}}= 0.812$. This is a 
%strong indication that our fit technique adequately represents 
%known charm contributions to the signal and the background.

In order to reproduce a background shape which may contain unsimulated backgrounds,
we took events which had at least one extra track consistent with the 
secondary vertex at vertex fit confidence levels between $30$\% and $90$\%.
In order to gauge the specific effect of non-simulated backgrounds on the final result, 
invariant mass histograms were formed from both the data and from a Monte Carlo in 
which we simulate all known charm decays except $D^+\to \overline{K}^0\mu^+\nu_\mu$
and $D^+\to \overline{K}^0\pi^0\mu^+\nu_\mu$. Using these shapes in addition to the
3 component fit significantly improves the fit quality. The $\chi^2/\textrm{DOF}$ for the
fit using the background from the data is acceptable ($\sim 1$ or less) at all
$\ell/\sigma_\ell$'s. There is now a component of the signal in the data background subtraction.
As a consequence, the signal and background are correlated in the fit, and the fit errors increase.
In the fit using the simulated background, we find that the $\chi^2/\textrm{DOF}$ 
is about 1 unit higher than the data represented background fit 
until $\ell/\sigma_\ell$'s above 11, where the fit quality becomes 
equivalent between the two representations. The difference in $\chi^2/\textrm{DOF}$ at 
low $\ell/\sigma_\ell$'s is likely due to non-charm or short-lived
charm decays, appearing primarily at low $K^0_S\mu$ mass according to the
results of the binning tests, which are not included in the simulation.
Even though we see a difference between the data and simulated background 
in $\chi^2/\textrm{DOF}$ at low $\ell/\sigma_\ell$, the results of the fits are 
in good agreement and stable above $\ell/\sigma_\ell=3$.

Our quoted result uses the fit to the $K^0_S\mu^+$ invariant mass
with the simulated background from higher multiplicity secondary vertices at 
$\ell/\sigma_\ell=13$. It is important to note that in all four of the fits described, 
the 3 component fit, the fit with the inclusive simulated background, and the 
fits with the background from higher multiplicity secondaries produce equivalent 
values above $\ell/\sigma_\ell=11$. This is likely due to
the background being dominated by $D\to K\pi\mu^+\nu_\mu$ decays at higher $\ell/\sigma_\ell$.

\begin{figure}[htb]
\begin{center}
\center{\epsfig{file=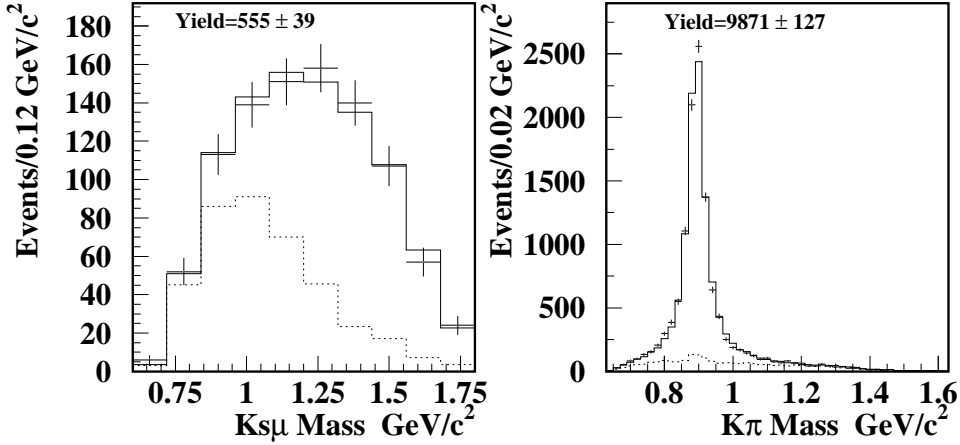,width=14cm}}
\caption{The fit to the $K^0_S\mu^+$ invariant  mass (left) and the $K^-\pi^+$ invariant  mass (right).
The fit to the data (error bars) is shown as a solid line, and the background described
in the text in shown as a dotted line. For both plots, the difference between the solid and dotted lines
represents the signal. Note the good fit to the ${K}^0_S\mu$ 
invariant mass and the cleanliness of the $K^-\pi^+$ invariant mass.}
\label{fig:fitfig}
\end{center}
\end{figure}

The fit to the data for both modes is presented in Fig. \ref{fig:fitfig}. We find 
$555\pm39~K^0_S\mu^+\nu_\mu$ decays and $9871\pm127~K^-\pi^+\mu^+\nu_\mu$ decays.  

%In the fit to the ${K}^0_S\mu$ invariant mass,
%the signal for the decay $D^+\to \overline{K}^0\mu^+\nu_\mu$, where the 
%$ \overline{K}^0$ has been reconstructed via the decay ${K}^0_S\to\pi^+\pi^-$, is 
%represented as the difference between the solid and dotted lines.
%In the fit to the $K^-\pi^+$ invariant mass,
%the signal for the decay $D^+\to ({K}\pi)\mu^+\nu_\mu$, where the ${K}\pi$ is reconstructed via ${K}^-\pi^+$,
%is represented as the difference between the solid and dotted lines. 

Thus, our ratio ${{\Gamma(D^+\to K\pi\mu^+\nu_\mu)}\over{\Gamma(D^+\to \overline{K}^0\mu^+\nu_\mu)}}$ becomes:

$${{1/2}\over{2/3}}~{{\# K^-\pi^+\mu^+\nu_\mu (\textrm{FIT})}\over {\# K^0_S\mu^+\nu_\mu (\textrm{FIT})}}~
{{\epsilon (K^0_S\mu^+\nu_\mu)}\over{\epsilon (K^-\pi^+\mu^+\nu_\mu)}}$$

The $2/3$ accounts for the probability that $\overline{K}^{*0}$ decays to $K^-\pi^+$ (see above) and the $1/2$
accounts for the probability that $K^0$ decays to $K^0_S$. The $K^0_S \to \pi^+\pi^-$ branching fraction
is accounted for in the Monte Carlo generation. The number of events determined from the fit to the data
for each mode is labeled (FIT), and the reconstruction efficiency for each mode determined using 
the Monte Carlo indicated by an $\epsilon$. In order to quote
the K* component as a separate result, we separate the resonant and non-resonant components using the
technique outlined in \cite{Link:2002wg}. Fully $95.0\pm0.5$\% of the $K^-\pi^+$ sample is proceeds
through a $\overline{K}^{*0}$. Thus we find: 
 
$${{\Gamma(D^+\to K\pi\mu^+\nu)}\over{\Gamma(D^+\to \overline{K}^0\mu^+\nu)}}=0.625\pm0.045$$

and, 

$${{\Gamma(D^+\to \overline{K}^{*0}\mu^+\nu)}\over{\Gamma(D^+\to \overline{K}^0\mu^+\nu)}}=0.594\pm0.043.$$

Our systematic tests of the result are outlined in the next section.

\section{Systematic Checks}

Our systematic uncertainty comes from known quantities which are not included in the fit
(such as the S-wave contribution estimate), unanticipated variations in the data not
accounted for in the simulation, and variations due to the fitting technique. 

To determine the amount of $D^+\to\overline{K}^{*0}\mu^+\nu_\mu$ contained in
our $D^+\to K^-\pi^+\mu^+\nu$ signal, we corrected our estimated yield of  
$D^+\to K^-\pi^+\mu^+\nu$ by $0.950\pm0.005$. We compute this fraction by integrating
over the $K^-\pi^+\mu^+\nu$ phase space the model intensity and parameters from  
reference \cite{Link:2002wg} with the S-wave amplitude set to zero and dividing this value by 
the same with the S-wave amplitude and phase set to the measured values. The 
uncertainty $(\pm0.005)$ is determined by varying the amplitude and phase by the errors indicated
in \cite{Link:2002wg} and noting the difference. 

In order to assess a systematic
uncertainty due to larger variations of $M_\textrmn{pole}$, we varied
$M_\textrmn{pole}$ between $1.86$ and $2.31~\textrm{GeV/c}^2$ in the Monte Carlo generation.
The resultant simulated signal histograms are then used to repeat the fit 
used to obtain the final result. The sample variance from the returned fit
results is retained as the systematic uncertainty due to $M_\textrmn{pole}$.
 To estimate the systematic uncertainty  due to different values of $f_-/f_+$, 
we repeated the fit at $M_\textrmn{pole}=2.11 ~\textrm{GeV/c}^2$ with $f_-/f_+=0.7$ and 
kept the difference as the error estimate.

In order to assess a systematic error to the measured ratio from unanticipated variations 
in the data not accounted for in the simulation, we placed a variety of pertinent
cuts on the data  and computed a sample variance from the returned values. 
The cuts described below are in addition to those previously applied. Unless
a particular meson or mode is mentioned, cuts are applied to both modes used
to calculate the ratio simultaneously.

To check for non-$K^0_S$ $\pi^+\pi^-$ backgrounds, we set the normalized 
$K^0_S$ mass cut to values of 3 and 1. To check for backgrounds where a 
neutral long-lived particle such as a $\Lambda$ is misidentified as a $K^0_S$,
we made a cut requiring that the difference in the magnitude of the $K^0_S$ 
candidate pion momenta be no greater than $70$\% of the sum. We also increased
the \v{C}erenkov requirement on both pions so that 
$W_\textrmn{obs}(K)-W_\textrmn{obs}(\pi)$ (pionicity) be greater than 0.0. Since 
this latter cut primarily removes low momentum $K^0_S$'s, 
it is an effective tool at examining the $K^0_S$ acceptance as well.

To check for unanticipated backgrounds and differences between the simulation and
the signal mode due to event topology, we investigated how the ratio behaves with
a variety of requirements on the detailed location of the reconstructed event in the
spectrometer. We required the $D^+$ vertex be located 
downstream of the first interaction target, 
downstream of the second interaction target, upstream of a trigger counter located 
near the main silicon tracking system, or upstream of a location roughly between 
the 2 downstream target silicon system planes. We also required the $K^0_S$ 
vertex be 2 cm downstream of the $D^+$ vertex, upstream of the main (last 12 planes) silicon system, 
or downstream of the target (1st 4 planes) silicon system \cite{Link:2002zg}.

To look for short lived or non-charm background, we increased the requirement to 3 standard deviations
that the $D^+$ vertex occur outside of target material, and we required $P_\textrmn{visible}>30 ~\textrm{GeV/c}^2$. As
a check for higher multiplicity decays feeding into the signal, we specified that the maximum
allowable confidence level that an additional track be consistent with the secondary be $1$\%.

In order to specifically look for decays feeding into the signal where a particle 
is misidentified as a muon, we chose cuts that should reduce the probability of 
contamination while leaving high efficiency for signal. We placed a cut on the muon 
requiring that the momentum measured by both magnets in the spectrometer be consistent. 
We increased the muon momentum cut to $>20~ \textrm{GeV/c}^2$. We also increased the 
requirement on the  muon identification confidence level to $15$\%.

To check for additional background in the $K\pi$ mass,
we increased the \v{C}erenkov likelihood difference cut on the kaon from $>2.0$ to $>4.0$,
and increased the cut on the mass difference used to cut out $D^*$'s to 0.25 and 0.30.

We assess a systematic uncertainty from these cut tests by computing a sample variance
of the returned values for ${{\Gamma(D^+\to K\pi\mu^+\nu)}\over{\Gamma(D^+\to \overline{K}^0\mu^+\nu)}}$.
The larger contributions to this systematic error estimate are listed in Table \ref{tab:sys}. 

In order to test the fit, we performed repeated tests where we Poisson fluctuated the bins
of both the data and fit histograms and performed repeated fits. Our tests indicate that
the $K_S^0\mu^+\nu_\mu$ yield error is underestimated by 6\% (due to finite Monte Carlo statistics), 
and the statistical error in the final result is boosted to account for this difference. 
We also looked at the standard deviation between the four different fits tried. Although
the fits agree remarkably well at $\ell/\sigma_\ell=13$, we felt a more conservative 
approach was to choose a value that was common to four returned ratios below and 
above the chosen $\ell/\sigma_\ell$. The systematic uncertainty estimated from the 
four fits is added in quadrature to the previously described estimates to assess a
total systematic uncertainty in the ratio (see Table 3).

\begin{table}[htb]
\caption{Larger contributions to the systematic uncertainty in the ratio ${{\Gamma(D^+\to
K\pi\mu^+\nu)}\over{\Gamma(D^+\to \overline{K}^0\mu^+\nu)}}$. Note that the uncertainty
associated with the non-S-wave $K\pi$ contribution applies only to the estimate of 
the $K^*$ fraction in the $K\pi$ signal.}
\label{tab:sys}
\begin{center}
\begin{tabular}{cccl} \hline 
Systematic Contribution & Value \\ \hline
Normalized $K^0_S$ Mass Cut             & 0.008 \\ 
Secondary Vertex Location               & 0.017 \\ 
$K^0_S$ Vertex Location                 & 0.013 \\  
Muon Magnet Consistency                 & 0.012 \\ 
Muon Momentum Cut                       & 0.008 \\ \hline
Total contributions from cut variations & 0.028\\ \hline
$M_\textrmn{pole}$ and $f_-/f_+$ variation & 0.015 \\ \hline
Contribution from fit variations        & 0.013\\ \hline
Total systematic uncertainty            & 0.034\\ \hline
S-wave Fraction ($K^*$ ratio only)      &  0.003 \\ \hline
\end{tabular}
\end{center}
\end{table}

\section{Summary and Conclusions }

Our result represents a substantial improvement over previous results for the 
ratio of the vector to pseudoscalar decay in the muon channel. We find 

$${{\Gamma(D^+\to K\pi\mu^+\nu)}\over{\Gamma(D^+\to \overline{K}^0\mu^+\nu)}}=
0.625\pm0.045\pm0.034$$

and
 
$${{\Gamma(D^+\to \overline{K}^{*0}\mu^+\nu)}\over{\Gamma(D^+\to \overline{K}^0\mu^+\nu)}}=
0.594\pm0.043\pm0.033.$$

Where the first error is statistical and the second error is the systematic uncertainty.
The $K^*/K$ result agrees with older results, but disagrees with a recent result from
the CLEO collaboration (with no corrections ($\sim 3$\% effect) made
for phase space and form factors). One possible source of this difference can be due to
the vector decay as is discussed in Reference \cite{Link:2002wg}. 
Another source is detailed below.  

By using our vector/pseudoscalar result and our measurement of the branching ratio 
${{\Gamma(D^+\to \overline{K}^{*0}\mu^+\nu)}\over{\Gamma(D^+\to K^-\pi^+\pi^+)}}$ \cite{Link:2002nj},
correcting for the updated values of the S-wave contribution \cite{Link:2002wg}, we determine:

$${{\Gamma(D^+\to \overline{K}^0\mu^+\nu)}\over{\Gamma(D^+\to K^-\pi^+\pi^+)}}=1.019\pm0.076
\pm0.065,$$
where we have added the statistical and systematic errors separately in quadrature.

Using the PDG \cite{Hagiwara:fs} values for the absolute branching fraction of the decay
$D^+\to K^-\pi^+\pi^+$: $(9.1\pm06)$\%, we calculate,

$$\textrm{B}(D^+\to \overline{K}^0\mu^+\nu)=(9.27\pm0.69\pm0.59\pm0.62)~\%.$$

The third error is due to uncertainty in the $D^+\to K^-\pi^+\pi^+$ branching fraction.
Our result is a substantial improvement over the present world average \cite{Hagiwara:fs} 
of $7.0^{+3.0}_{-2.0}$\%. Besides the difference between the $D^+$ and $D^0$
semi-electronic rates mentioned in the introduction,
there are other reasons to believe that the PDG 
value for the $D^+$ semi-electronic mode $6.5\pm0.9$\% is low.

We can compare the $D^0$ and $D^+$ semi-muonic rates as we did for the semi-electronic 
rates in the introduction. We find good agreement with $\Gamma(D^+\to \overline{K}^0\mu^+\nu_\mu)-
\Gamma(D^0\to K^-\mu^+\nu_\mu)=11\pm11~\textrm{ns}^{-1}$ (where no correction for the
difference in phase space between $D^+$ and $D^0$ has been made). 

It is likely that the semi-muonic rate is lower than the semi-electronic rate by
a few percent, and this is consistent with what is measured for the isospin conjugate 
decay of the $D^0$ using values from \cite{Hagiwara:fs}.
One sees ${{\Gamma(D^0\to K^-e^+\nu_e)/\Gamma(D^0\to K^-\pi^+)}
\over{\Gamma(D^0\to K^-\mu^+\nu_\mu)/\Gamma(D^0\to K^-\pi^+)}}=1.12\pm0.07$,
for the $D^0$ but ${{\Gamma(D^+\to K^0e^+\nu_e)/\Gamma(D^+\to K^-\pi^+\pi^+)}
\over{\Gamma(D^+\to \overline{K}^0\mu^+\nu_\mu)/\Gamma(D^+\to K^-\pi^+\pi^+)}}=0.72\pm0.11$
for the $D^+$ using the result in this paper. The PDG estimates that this
ratio should be around 1.03\cite{Hagiwara:fs}. This is reasonable since
a very large positive ratio for $f_-/f_+$, which increases the semi-muonic rate, is
not likely given the E687 \cite{Frabetti:1995xq} result for $f_-/f_+$, and
radiative corrections for the semi-electronic mode, which can lower 
the measured semi-electronic rate, are expected to be small. It is also unlikely
that any such large, unanticipated corrections apply to the 
$D^+$ semi-electronic mode exclusively.

Finally, it is 
interesting to note that the sum of the rates measured by FOCUS, corrected as suggested in the 
PDG \cite{Hagiwara:fs} to estimate the semi-electronic modes, is 
$\textrm{B}(D^+\to ((1.05){K}\pi+(1.03)K^0)\mu^+\nu_\mu)=14.9\pm1.2$\%.
This is closer to the current world average inclusive electronic rate $D^+\to
e^+$(anything)$=17.2\pm1.9$\%  
than 
$\textrm{B}(D^+\to ((3/2){K^-}\pi^++K^0)\mu^+\nu_\mu)=12.9^{+1.6}_{-1.4}$\% (both from \cite{Hagiwara:fs}), 
suggesting that a large, previously unseen, semileptonic decay mode for the $D^+$ is unlikely.

\section*{Acknowledgments}

We wish to acknowledge the assistance of the staffs of Fermi
National Accelerator Laboratory, the INFN of Italy, and the physics
departments of the collaborating institutions. This research was 
supported in part by the U.~S.
National Science Foundation, the U.~S. Department of Energy, the
Italian Istituto Nazionale di F{i}sica Nucleare and Ministero
dell'Universit\`a e della Ricerca Scientifica e Tecnologica, 
the Brazilian Conselho Nacional de
Desenvolvimento Cient\'{\i}fico e Tecnol\'ogico, CONACyT-M\'exico,
the Korean Ministry of Education, and the Korean Science and 
Engineering Foundation.

\afterpage{\clearpage}

\clearpage


\begin{thebibliography}{00}

%%
%%  bibliographic items can be constructed using the LaTeX format in SPIRES:
%%    see    http://www.slac.stanford.edu/spires/hep/latex.html
%%  SPIRES will also supply the CITATION line information; please include it.
%%

%\cite{Link:2002wg}
\bibitem{Link:2002wg}
J.~M.~Link {\it et al.}  [FOCUS Collaboration],
%``New measurements of the D+ $\to$ anti-K*0 mu+ nu form factor ratios,''
Phys.\ Lett.\ B {\bf 544}, 89 (2002).
%[arXiv:hep-ex/0207049].
%%CITATION = HEP-EX 0207049;%%

%\cite{Hagiwara:fs}
\bibitem{Hagiwara:fs}
D.E. Groom {\it et al.}  [Particle Data Group Collaboration],
%``Review Of Particle Physics,''
Phys. \ Rev. \ D {\bf} 66, 010001 (2002) and 2003 partial update 
for edition 2004 (URL:http//pdg.lbl.gov).
%%CITATION = PHRVA,D66,010001;%%

%\cite{Wirbel:1985ji}
\bibitem{Wirbel:1985ji}
M.~Wirbel, B.~Stech and M.~Bauer,
%``Exclusive Semileptonic Decays Of Heavy Mesons,''
Z.\ Phys.\ C {\bf 29}, 637 (1985).
%%CITATION = ZEPYA,C29,637;%%
%WSB(85) Z.Phys. C29 (1985) 637 

%\cite{Brandenburg:2002eu}
\bibitem{Brandenburg:2002eu}
G.~Brandenburg {\it et al.}  [CLEO Collaboration],
%``Measurement of the D+ $\to$ anti-K*0 l+ nu/l branching fraction,''
Phys.\ Rev.\ Lett.\  {\bf 89}, 222001 (2002).
%[arXiv:hep-ex/0203030].
%%CITATION = HEP-EX 0203030;%%


%\cite{Crawford:1991zd}
\bibitem{Crawford:1991zd}
G.~D.~Crawford {\it et al.}  [CLEO Collaboration],
%``Measurement of the ratio B (D0 $\to$ K*- e+ electron-neutrino) / B (D0 $\to$
%K- e+ electron-neutrino),''
Phys.\ Rev.\ D {\bf 44}, 3394 (1991).
%%CITATION = PHRVA,D44,3394;%%

%\cite{Bean:zv}
\bibitem{Bean:zv}
A.~Bean {\it et al.}  [CLEO Collaboration],
%``Measurement Of Exclusive Semileptonic Decays Of D Mesons,''
Phys.\ Lett.\ B {\bf 317}, 647 (1993).
%%CITATION = PHLTA,B317,647;%%

%\cite{Frabetti:1990au}
\bibitem{Frabetti:1990au}
P.~L.~Frabetti {\it et al.}  [E687 Collaboration],
%``Description And Performance Of The Fermilab E687 Spectrometer,''
Nucl.\ Instrum.\ Meth.\ A {\bf 320}, 519 (1992).
%%CITATION = NUIMA,A320,519;%%

%\cite{Link:2002zg}
\bibitem{Link:2002zg}
J.~M.~Link {\it et al.}  [FOCUS Collaboration],
%``The target silicon detector for the FOCUS spectrometer,''
Nucl.\ Instrum.\ Meth.\ A {\bf 516}, 364 (2004).
%%CITATION = HEP-EX 0204023;%%

%\cite{Link:2002ev}
\bibitem{Link:2002ev}
J.~M.~Link {\it et al.}  [FOCUS Collaboration],
%``Evidence for new interference phenomena in the decay  D+ $\to$ K- pi+ mu+ nu,''
Phys.\ Lett.\ B {\bf 535}, 43 (2002).
%[arXiv:hep-ex/0203031].

%\cite{Link:2002nj}
\bibitem{Link:2002nj}
J.~M.~Link {\it et al.}  [Focus Collaboration],
 %``New measurements of the Gamma(D+ $\to$ anti-K*0 mu+ nu)/Gamma(D+ $\to$  K-
%pi+ pi+) and Gamma(D/s+ $\to$ Phi mu+ nu)/Gamma(D/s+ $\to$ Phi pi+)  branching
%ratios,''
Phys.\ Lett.\ B {\bf 541}, 243 (2002).
%[arXiv:hep-ex/0206056].
%%CITATION = HEP-EX 0206056;%%

%\cite{Frabetti:1992bn}
\bibitem{Frabetti:1992bn}
P.~L.~Frabetti {\it et al.},
%``A Wide band photon beam at the Fermilab Tevatron to study heavy flavors,''
Nucl.\ Instrum.\ Meth.\ A {\bf 329}, 62 (1993).
%%CITATION = NUIMA,A329,62;%%


%\cite{Link:2001pg}
\bibitem{Link:2001pg}
J.~M.~Link {\it et al.}  [FOCUS Collaboration],
%``Cerenkov particle identification in FOCUS,''
Nucl.\ Instrum.\ Meth.\ A {\bf 484}, 270 (2002).
%[arXiv:hep-ex/0108011].
%%CITATION = HEP-EX 0108011;%%



%\cite{Anjos:1988ue}
\bibitem{Anjos:1988ue}
J.~C.~Anjos {\it et al.}  [Tagged Photon Spectrometer Collaboration],
%``A Study Of The Semileptonic Decay Mode D0 $\to$ K- E+ Electron-Neutrino,''
Phys.\ Rev.\ Lett.\  {\bf 62}, 1587 (1989).
%%CITATION = PRLTA,62,1587;%%

%\cite{Frabetti:1993vz}
\bibitem{Frabetti:1993vz}
P.~L.~Frabetti {\it et al.}  [E687 Collaboration],
%``Study of D0 $\to$ K- mu+ neutrino in high-energy photoproduction,''
Phys.\ Lett.\ B {\bf 315}, 203 (1993).
%%CITATION = PHLTA,B315,203;%%

%\cite{Frabetti:1995xq}
\bibitem{Frabetti:1995xq}
P.~L.~Frabetti {\it et al.}  [E687 Collaboration],
%``Analysis of the decay mode D0 $\to$ K- mu+ muon-neutrino,''
Phys.\ Lett.\ B {\bf 364}, 127 (1995).
%%CITATION = PHLTA,B364,127;%%

%\cite{PDG00:2000sj}
\bibitem{PDG00:2000sj}
S.~J.~Lee,
Eur.\ Phys.\ J.\ C {\bf 15}, 1 (2000).
%%CITATION = EPHJA,C15,1;%%

%\cite{Link:2001zj}
\bibitem{Link:2001zj}
J.~M.~Link {\it et al.}  [FOCUS Collaboration],
 %``Search for CP violation in the decays D+ $\to$ K(S) pi+ and D+ $\to$ K(S)
%K+,''
Phys.\ Rev.\ Lett.\  {\bf 88}, 041602 (2002)
[Erratum-ibid.\  {\bf 88}, 159903 (2002)].
%[arXiv:hep-ex/0109022].
%%CITATION = HEP-EX 0109022;%%

%\cite{PDG00:2000sj}
\bibitem{Cleo:Yuichi} Y. Kubota, private communication.
%%CITATION = EPHJA,C15,1;%%



%\cite{Jaus:au}
\bibitem{Jaus:au}
W.~Jaus,
%``Semileptonic Decays Of B And D Mesons In The Light Front Formalism,''
Phys.\ Rev.\ D {\bf 41}, 3394 (1990).
%%CITATION = PHRVA,D41,3394;%%
%LFR(90) Phys.Rev.D41:3394,1990 

%\cite{Ball:bs}
\bibitem{Ball:bs}
P.~Ball, V.~M.~Braun and H.~G.~Dosch,
%``Form-Factors Of Semileptonic D Decays From QCD Sum Rules,''
Phys.\ Rev.\ D {\bf 44}, 3567 (1991).
%%CITATION = PHRVA,D44,3567;%%
%SUMR(91) Phys.Rev.D44:3567-3581,1991 

%\cite{Lubicz:1991bi}
\bibitem{Lubicz:1991bi}
V.~Lubicz, G.~Martinelli, M.~S.~McCarthy and C.~T.~Sachrajda,
%``Semileptonic decays of D mesons in a lattice QCD,''
Phys.\ Lett.\ B {\bf 274}, 415 (1992).
%%CITATION = PHLTA,B274,415;%%
%LAT(92)Phys.Lett.B274:415-420,1992  

%\cite{Casalbuoni:1992dx}
\bibitem{Casalbuoni:1992dx}
R.~Casalbuoni, A.~Deandrea, N.~Di Bartolomeo, R.~Gatto, F.~Feruglio and G.~Nardulli,
%``Effective Lagrangian for heavy and light mesons: Semileptonic decays,''
Phys.\ Lett.\ B {\bf 299}, 139 (1993).
%[arXiv:hep-ph/9211248].
%%CITATION = HEP-PH 9211248;%%
%LAGR(93)Phys.Lett.B299:139-150,1993 hep-ph/9211248

%\cite{Sadzikowski:iy}
\bibitem{Sadzikowski:iy}
M.~Sadzikowski,
%``Semileptonic Decays Of Heavy To Light Mesons From An Mit Bag Model,''
Z.\ Phys.\ C {\bf 67}, 129 (1995).
%[arXiv:hep-ph/9407263].
%%CITATION = HEP-PH 9407263;%%
%MITBAG(94) hep-ph 9407263

%\cite{Abada:1993dh}
\bibitem{Abada:1993dh}
A.~Abada {\it et al.},
%``Semileptonic decays of heavy flavors on a fine grained lattice,''
Nucl.\ Phys.\ B {\bf 416}, 675 (1994).
%[arXiv:hep-lat/9308007].
%%CITATION = HEP-LAT 9308007;%%
%LAT(94)Published in Nucl.Phys.B416:675-698,1994 e-Print Archive: hep-lat/9308007

%\cite{Scora:1995ty}
\bibitem{Scora:1995ty}
D.~Scora and N.~Isgur,
%``Semileptonic meson decays in the quark model: An update,''
Phys.\ Rev.\ D {\bf 52}, 2783 (1995).
%[arXiv:hep-ph/9503486].
%%CITATION = HEP-PH 9503486;%%
%ISGW2(95) hep-ph/9503486 

%\cite{Faustov:xe}
\bibitem{Faustov:xe}
R.~N.~Faustov, V.~O.~Galkin and A.~Y.~Mishurov,
%``Relativistic Description Of Exclusive Semileptonic Decays Of Heavy Mesons,''
Phys.\ Rev.\ D {\bf 53}, 1391 (1996).
%%CITATION = PHRVA,D53,1391;%%
%QPT(96) Phys. Rev. D 53, 1391-1402 (1996)

%\cite{Jaus:np}
\bibitem{Jaus:np}
W.~Jaus,
%``Semileptonic, Radiative, And Pionic Decays Of B, B* And D, D* Mesons,''
Phys.\ Rev.\ D {\bf 53}, 1349 (1996)
[Erratum-ibid.\ D {\bf 54}, 5904 (1996)].
%%CITATION = PHRVA,D53,1349;%%
%RQM(96) Phys. Rev. D 53, 1349-1365 (1996)

%\cite{Melikhov:1996pr}
\bibitem{Melikhov:1996pr}
D.~Melikhov,
%``Exclusive semileptonic decays of heavy mesons in quark model,''
Phys.\ Lett.\ B {\bf 394}, 385 (1997).
%[arXiv:hep-ph/9611364].
%%CITATION = HEP-PH 9611364;%%
%QM(97) Phys.Lett.B394:385-394,1997  hep-ph/9611364

%\cite{Demchuk:1997uz}
\bibitem{Demchuk:1997uz}
N.~B.~Demchuk, P.~Y.~Kulikov, I.~M.~Narodetsky and P.~J.~O'Donnell,
%``Light-front model for exclusive semileptonic B and D decays,''
Phys.\ Atom.\ Nucl.\  {\bf 60}, 1292 (1997)
[Yad.\ Fiz.\  {\bf 60N8}, 1429 (1997)].
%[arXiv:hep-ph/9701388].
%%CITATION = HEP-PH 9701388;%%
%LFM(97) hep-ph 9701388

%\cite{Yang:ar}
\bibitem{Yang:ar}
K.~C.~Yang and W.~Y.~P.~Hwang,
 %``The QCD Sum Rule Approach For The Semileptonic Decay Of The D Or B  Meson
%Into A Light Meson And Leptons,''
Z.\ Phys.\ C {\bf 73}, 275 (1997).
%%CITATION = ZEPYA,C73,275;%%
%SR(97) Z. Phys C 73, 275-292 (1997)

%\cite{Melikhov:2000yu}
\bibitem{Melikhov:2000yu}
D.~Melikhov and B.~Stech,
%``Weak form factors for heavy meson decays: An update,''
Phys.\ Rev.\ D {\bf 62}, 014006 (2000).
%[arXiv:hep-ph/0001113].
%%CITATION = HEP-PH 0001113;%%
%QM(00)  hep-ph 0001113

%\cite{Merten:2001er}
\bibitem{Merten:2001er}
D.~Merten, R.~Ricken, M.~Koll, B.~Metsch and H.~Petry,
%``Weak decays of heavy mesons in a covariant quark model,''
Eur.\ Phys.\ J.\ A {\bf 13}, 477 (2002).
%[arXiv:hep-ph/0104029].
%%CITATION = HEP-PH 0104029;%%
%COVQ(2001) hep-ph 0104029 

%\cite{Nasriddinov:qh}
\bibitem{Nasriddinov:qh}
K.~R.~Nasriddinov, B.~N.~Kuranov, G.~G.~Takhtamyshev and T.~A.~Merkulova,
%``Pole Contributions In Semileptonic Decays Of D Mesons,''
Phys.\ Atom.\ Nucl.\  {\bf 64}, 1323 (2001)
[Yad.\ Fiz.\  {\bf 64}, 1399 (2001)].
%%CITATION = PANUE,64,1323;%%
%CVLFD(2001) Physics of atomic nuclei v64 no4

%\cite{Melikhov:2001zv}
\bibitem{Melikhov:2001zv}
D.~Melikhov,
 %``Dispersion approach to quark-binding effects in weak decays of heavy
%mesons,''
Eur.\ Phys.\ J.\ directC {\bf 4}, 2 (2002).
%[arXiv:hep-ph/0110087].
%%CITATION = HEP-PH 0110087;%%

%DISP(01) hep-ph/0110087

%\cite{Kondratyuk:2000ka}
\bibitem{Kondratyuk:2000ka}
L.~A.~Kondratyuk and D.~V.~Chekin,
 %``Transition form factors and widths of semileptonic decays  of B and D mesons
%in the covariant light-front dynamics. (In Russian),''
Phys.\ Atom.\ Nucl.\  {\bf 64}, 727 (2001)
[Yad.\ Fiz.\  {\bf 64}, 786 (2001)].
%%CITATION = PANUE,64,727;%%
%PCL(01) Physics of atomic nuclei v64 no7


%\cite{Wang:2002zb}
\bibitem{Wang:2002zb}
W.~Y.~Wang, Y.~L.~Wu and M.~Zhong,
 %``Heavy to light meson exclusive semileptonic decays in effective field
%theory of heavy quarks,''
Phys.\ Rev.\ D {\bf 67}, 014024 (2003).
%[arXiv:hep-ph/0205157].
%%CITATION = HEP-PH 0205157;%%
%EFT(2002) hep-ph 0205157












\end{thebibliography}
\end{document}